\documentclass[aps,pra,twocolumn,showpacs,superscriptaddress,groupedaddress]{revtex4-1}

\usepackage{amsmath}
\usepackage{amsthm}
\usepackage{newlfont}
\usepackage{graphicx}
\usepackage{epstopdf}
\usepackage{appendix}
\usepackage{hyperref}
\usepackage{textcomp}
\usepackage{appendix}
\usepackage{multirow}	
\usepackage{hyperref}
\usepackage{color}
\usepackage{amssymb}
\usepackage{epsfig}
\usepackage{bm}
\usepackage[american]{babel}
\usepackage{braket}

\hypersetup{colorlinks=true,linkcolor=blue,citecolor=blue,filecolor=blue,urlcolor=blue,pdfstartview=FitH}
	
\begin{document}
\title{Modulation instability associated nonlinear dynamics of spin-orbit coupled Bose-Einstein condensates}
\author{Thudiyangal Mithun$^1$, and Kenichi Kasamatsu$^2$}
\affiliation{$^1$Center for Theoretical Physics of Complex Systems, Institute for Basic Science, Daejeon, Korea}
%\email{kenichi@phys.kindai.ac.jp}
\affiliation{$^2$Department of Physics, Kindai University, Higashi-Osaka, Osaka 577-8502, Japan}

\begin{abstract}
We study pattern-forming nonlinear dynamics starting from a continuous wave state of quasi-one-dimensional two-component 
Bose-Einstein condensates with synthetic spin-orbit coupling induced by Raman lasers. 
Modulation instability can occur even when the miscibility condition due to the interatomic interactions 
is satisfied. We find that the initial stage of the nonlinear development is consistent with the prediction of modulation 
instability, where the two primary and secondary instability bands lead to the spontaneous growth of the modulation 
and the subsequent complicated dynamics of pattern formation. At later stages of the evolution, the wave functions 
undergo clear separation in the momentum space, reflected in the dispersion of the single particle Hamiltonian.

\emph{T. Mithun dedicates this article to the memory of beloved guide Prof. K. Porsezian, who passed away on August 11, 2018.}
\end{abstract}
\maketitle
\pagestyle{myheadings}

\section{Introduction}{\label{intro}}
Modulation instability (MI) is one of the most fundamental process of nonlinear wave dynamics in various systems \cite{zakharov2009modulation}.
The instability undergoes a spontaneous growth from small-amplitude modulated waves and leads to 
stable large-amplitude localized waves, typically solitary waves, as a result of 
interplay between the intrinsic nonlinearity and the dispersion. 
The MI has been studied in the nonlinear dynamics 
of matter waves, corresponding to Bose-Einstein condensates (BECs) in ultracold atomic 
gases \cite{kevrekidis2007emergent}. The dynamics of the matter waves is described by the nonlinear Schr\"{o}dinger equation, also 
known as the Gross-Pitaevskii (GP) equation, where the nonlinearity is associated with 
the interatomic collisions. The remarkable feature of this system is that 
the dispersion as well as nonlinearity can be controlled experimentally in a well controlled manner \cite{morsch2006dynamics,chin2010feshbach}. 
In addition, there is diverse richness as the nonlinear wave system; for 
example, we can consider the system of multicomponent order parameters 
with various linear and nonlinear couplings between them. 

The MI in scalar BECs has been firstly discussed in the context of the formation 
of bright soliton trains \cite{theocharis2003modulational,salasnich2003modulational,carr2004spontaneous}. 
This has been experimentally demonstrated by tuning the interaction of condensed 
atoms from repulsive to attractive \cite{strecker2002formation,nguyen2017formation,everitt2017observation}. 
For a uniform scalar condensate, the MI is possible only for the attractive nonlinearity. 
Recently, the argument of MI has been discussed for the 
BEC with long-range interactions \cite{ferrier2018onset}, where 
the MI is related to the formation of quantum droplets as a 
result of the beyond mean-field corrections \cite{ferrier2016observation,schmitt2016self}. 
Also, the MI has been extended to the system of multicomponent BECs. 
The presence of the intercomponent interaction induces the MI even for the 
condensates with repulsive nonlinearity \cite{goldstein1997quasiparticle}. 
When the intercomponent repulsion is stronger than the intracomponent one, 
the MI induces the formation of multiple domains \cite{kasamatsu2004multiple,kasamatsu2006modulation,PhysRevA.85.043602,vidanovic2013spin,eto2016nonequilibrium}. 
Recent theoretical analysis has revealed that the formation and subsequent coalesce 
dynamics of condensate domains are governed by the universal scaling law 
\cite{sabbatini2011phase,de2014quenched,hofmann2014coarsening,takeuchi2015phase}. 

The recent papers on MI has shown that two-component BECs 
with spin-orbit coupling (SOC) are always subject to the MI 
for arbitrary choice of the nonlinearities \cite{Bhat:2015, Bhuvaneswari:2016,Congy:2016}.
Thus, the SOC extends the parameter region of the MI from that of the conventional 
two-component BECs. The SOC can be synthesized by 
the Raman laser coupling scheme between internal states of cold atoms \cite{lin2011spin}. 
The static and dynamical properties of the BECs with Raman-laser induced SOC have been studied in 
many papers 
\cite{lin2011spin,ho2011bose,li2012quantum,chen2017collective,martone2012anisotropic,zhang2012collective,zheng2013properties,achilleos2013matter,kasamatsu2015dynamics}, but the strong nonlinear evolution caused by MI has not been studied so much. 
Recently, Ye \textit{et al}. studied the domain formation through the parameter quench from the mixed phase to the 
plane-wave phase of the BEC with the SOC \cite{ye2018universal}. 

The objective of the present work is the analysis of nonlinear dynamics caused 
by MI in the effectively one-dimensional (1D) BEC with synthetic SOC by the numerical 
simulations of the GP equation. 
Following the MI analysis by Bhat \textit{et al} \cite{Bhat:2015}, we study the nonlinear 
dynamics starting from continuous wave (cw) states of miscible two-component BECs. 
By suddenly turning on the synthetic SOC, the system is modulationally unstable and 
there appears a pulse-like structure. The initial stage of the evolution is consistent 
with the prediction of the MI analysis, where the dynamically unstable modulations grow spontaneously from 
primary and secondary instability bands in small-$k$ and large-$k$ regions.
The subsequent nonlinear evolutions exhibit complicated dynamics of pattern formation in a 
real space, while clear separation of the wave functions of the two component BECs in a momentum 
space is observed due to the effect of the SOC.  Our results reveals the richness in the complex dynamics exhibiting by the SOC BECs, alongside of the recent experimental and theoretical observation of the spin dynamics and the dynamical instabilities \cite{Burdick:2016, Zhu:2016,Khamehchi:2017} .

The paper is organized as follows. In Sec.~\ref{MIofBEC}, we introduce the basic formulation 
of the problem and briefly review the MI in BECs with the Raman-induced SOC. 
Section~\ref{nonlinearMI} presents the results of the series of numerical simulations of the MI induced nonlinear dynamics. 
In Sec.~\ref{conclusion}, we devote to the conclusion. 
  
\section{Modulation instability of BECs with synthetic SOC}\label{MIofBEC}
In this section, we first introduce the basic formulation of BECs with a synthetic SOC in our problem. 
Next, we briefly review the MI analysis done by Bhat \textit{et al}. \cite{Bhat:2015} and 
specify the parameter region to see the MI-induced nonlinear dynamics. 
%-----------------
\subsection{Model}\label{model}
We consider quasi-1D two-component (psudospin-1/2) BECs 
with the combined Rashba-Dresselhaus SOC induced by the Raman lasers \cite{lin2011spin}. 
The single-particle hamiltonian with the synthetic SOC in a quasi-momentum frame has the $2\times 2$ matrix structure \cite{Zhu:2016} 
\begin{equation}
h_0 = \frac{1}{2m} \left(p_x \sigma_0- \hbar k_\text{R} \sigma_z \right)^2 
+ \frac{\hbar \delta}{2} \sigma_z+ \frac{\hbar\Omega_\text{R}}{2} \sigma_x + V_\text{tr} \sigma_0. 
\label{singleho}
\end{equation}
Here, $m$ is the atomic mass, $p_x = -i \hbar \partial_x$ the quasi-momentum operator 
along the $x$-direction, $\sigma_r$ for $r=x,y,z$ is one 
of the Pauli matrices and $\sigma_0$ is the unit matrix. The quasi-momentum 
is related with the real momentum $p_x'$ as $p_x' \sigma_0=p_x \sigma_0- \hbar k_\text{R} \sigma_z$.
The trapping potential is assumed to be a harmonic form 
$V_\text{tr} = m \omega^2 x^2 / 2$. 
The SOC is characterized by three parameters, $k_\text{R}$,  $\Omega_\text{R}$ and $\delta$ under experimental 
control \cite{lin2011spin}, where $k_\text{R}$ is the wavenumber of the Raman laser which couples the two atomic 
hyperfine states, $\Omega_\text{R}$ is the Rabi frequency determined by 
the intensity of the Raman laser, and $\delta$ is the detuning. 
For simplicity, the detuning $\delta$ is set to be zero. 
The kinetic energy term has a uniform synthetic gauge field 
$ -\hbar k_\text{R} \sigma_z$ proportional 
to the spin matrix $\sigma_z$, which represents the 1D SOC whose magnitude can be controlled by $k_\text{R}$.

The GP energy functional including the single-particle hamiltonian 
of Eq.~(\ref{singleho}) and the atom-atom interactions is given by
\begin{align}
E = \int dx \left( \Psi^\dagger h_0 \Psi + 
\frac{1}{2}  \sum_{j=1,2} u_{j}|\Psi_j|^4 + u_{12} |\Psi_1|^2 |\Psi_2|^2 \right). \label{EGPSO}
\end{align} 
The order parameters are represented by the 2-component spinor $\Psi = (\Psi_1,\Psi_2)^{T}$. 
The parameters $u_j$ and $u_{12}$ are the coupling constants adjusted to the quasi-1D description 
by incorporating the length scale along the tightly confined direction \cite{kasamatsu2006modulation}, 
being proportional to the $s$-wave scattering lengths of atoms. 
We have set the same intracomponent coupling constant as $u_1=u_2=u$ for simplicity.  
When we take the energy scale by the recoil energy $E_\text{R} = \hbar^2k_\text{R}^2/m$ and 
the length scale by $k_\text{R}^{-1}$, Eq.~\eqref{singleho} is scaled as
\begin{align}
\tilde{h}_0 = \frac{1}{2} \left(- i \frac{\partial}{\partial \tilde{x}} \sigma_0 - \tilde{\gamma} \sigma_z \right)^2  
+ \tilde{\Gamma} \sigma_x + \tilde{V}_{\mathrm{tr}} \sigma_0,  \label{singleph}
\end{align}
where the dimensionless quantities are represented by symbols with tildes. 
In our unit, although the parameter $\tilde{\gamma}$ is kept unity, we leave this notation 
because this parameter is used as a quench parameter to induce the MI. 
The Rabi frequency is written as $\tilde{\Gamma} = \hbar \Omega_\text{R} / (2 E_R)$ and
the trap potential as $\tilde{V}_\text{tr} = \lambda^2 \tilde{x}^2 / 2$ with 
the coefficient $\lambda = (a_\text{ho} k_\text{R})^{-2}$, where 
$a_\text{ho} = \sqrt{\hbar/(m\omega)}$ is the harmonic oscillator length. 
Following a usual experimental condition, we use $\lambda = 0.02$ in the following calculation.  
The normalization of the wave function is given by the total particle number in the 1D system 
$N = \int dx \Psi^{\dagger} \Psi =  \int d x (|\Psi_1|^2 + |\Psi_2|^2) = N_1 + N_2$. 
By replacing the wave function as $\Psi = \sqrt{N k_\text{R}} \tilde{\psi}$, 
we have $\int d \tilde{x} \tilde{\psi}^{\dagger} \tilde{\psi} = \int d \tilde{x} ( |\tilde{\psi}_1|^2 +  |\tilde{\psi}_2|^2) =1$ and 
define the dimensionless coupling strengths as $\tilde{g} = m N u / (\hbar^2 k_\text{R})$ and $\tilde{g}_{12} = m N u_{12} / (\hbar^2 k_\text{R})$. 
The time-dependent GP equations derived from Eq.~(\ref{EGPSO}) 
can be written as  
\begin{align}
i \frac{\partial \psi_1}{\partial t} = \left( -\frac{\partial_x^2}{2}  - i \gamma \partial_x + V_\text{tr} + g |\psi_1|^2 + g_{12} |\psi_2|^2 \right) \psi_1  \nonumber \\ 
+ \Gamma \psi_2,   \label{tdgp1} \\
i \frac{\partial \psi_2}{\partial t} = \left( -\frac{\partial_x^2}{2}  + i \gamma \partial_x + V_\text{tr} + g |\psi_2|^2 + g_{12} |\psi_1|^2 \right) \psi_2  \nonumber \\ 
+ \Gamma \psi_1,    \label{tdgp2}
\end{align}
 where tildes are omitted in the notation. The unit of time is taken as $\hbar/E_\text{R}$.
\subsection{Ground States}
   %-----------------------------------------
%      \begin{figure}[!htbp] 
      % \vspace{0pt}
%  \includegraphics[width=8.5cm,scale=1]{ground}%\hspace{-0.8em}
%     \vspace{-2.0pt}
% \caption{\label{Fig3}{\footnotesize (Color online) Density profile of ground state of Eqs. \ref{tdgp1}  and \ref{tdgp2} for $\Gamma=\gamma=0$. }}
%%      \vspace{-15pt}
%\label{ap:ground}
 % \end{figure}
 \begin{figure}[!htbp] 
  \includegraphics[width=4.2cm,scale=1]{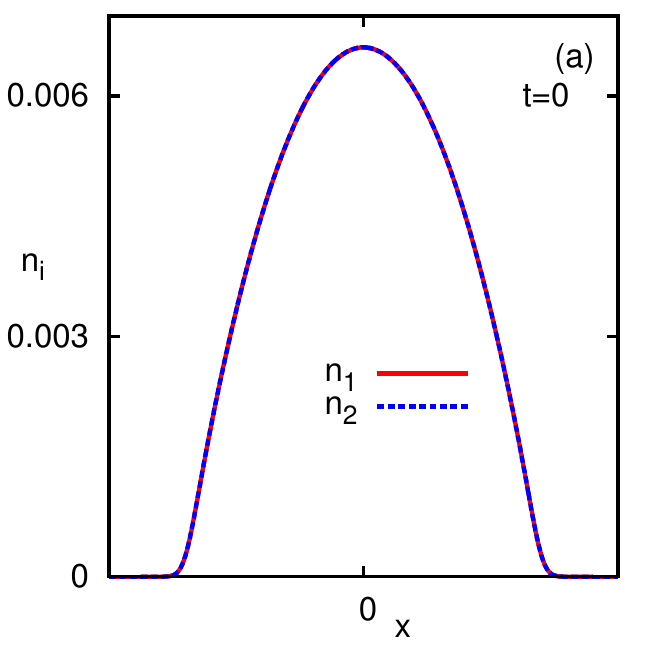}%\hspace{-0.8em}
  \includegraphics[width=4.2cm,scale=1]{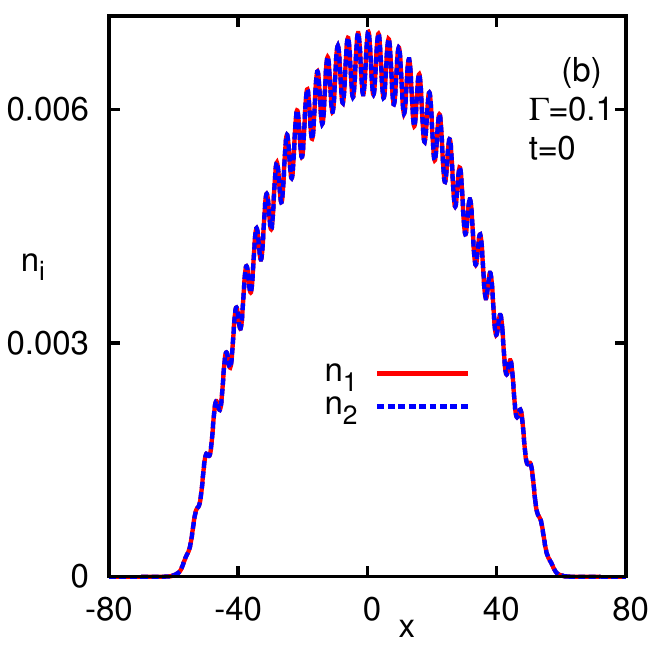} \\
  \includegraphics[width=4.2cm,scale=1]{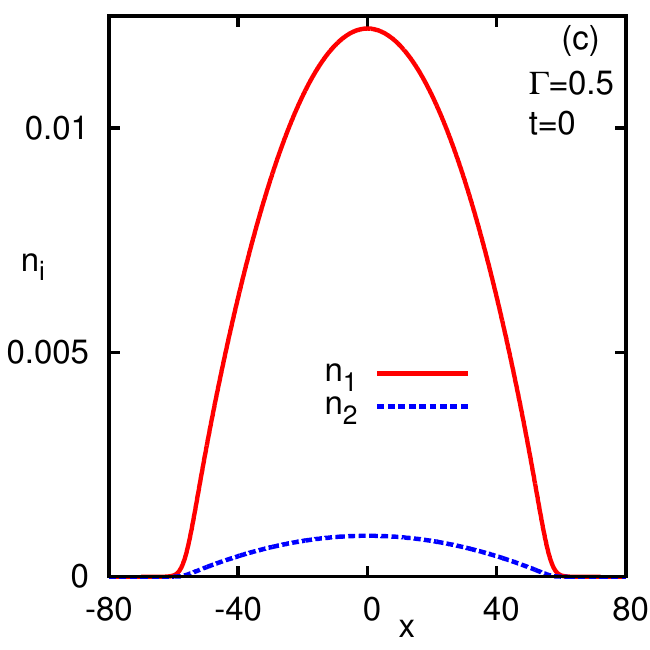} 
  \includegraphics[width=4.2cm,scale=1]{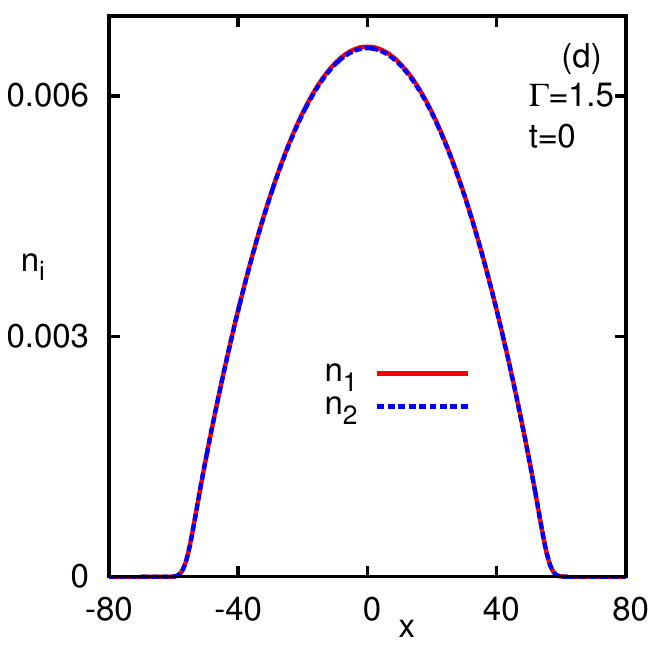}%\hspace{-0.8em}
   \vspace{-2.0pt}
\caption{\label{Fig1} {\footnotesize(Color online) Density profiles of the ground state of Eqs. \eqref{tdgp1} and \eqref{tdgp2} for the coupling constants $g=50$ and $g_{12}=0.95g$, 
and the trap frequency $\lambda=0.02$. Figure (a) represents the profile without the SOC, $\Gamma=\gamma=0$, which corresponds to the initial states of the 
time evolution shown below.  Figures (b) represents the ground state for $\Gamma=0.1$, and (c)  for $\Gamma=0.5$, and (d)  for $\Gamma=1.5$, 
corresponding to the stripe phase ($0<\Gamma \lesssim 0.2$), plane-wave phase ($0.2\lesssim \Gamma \lesssim 1$) and mixed phase ($1 \lesssim \Gamma$), respectively with $\gamma=1$. 
}}
  \end{figure}
  %-----------------------------------------
It is well known that two-component BECs without SOC have two types of the ground state phases characterized by the miscible or immiscible 
density profile \cite{ao1998binary,trippenbach2000structure}. This miscible-immiscible transition is associated with the relation of the coupling constant; when $g > g_{12}$ ($g < g_{12}$) 
the system is miscible (immiscible). This condition is deeply related to the MI in conventional two-component BECs \cite{goldstein1997quasiparticle,kasamatsu2004multiple,kasamatsu2006modulation}. 
When, the SOC is present, for a given strength of inter and intracomponent interaction, there exist three different ground state phases 
depending on the Rabi frequency $\Gamma$ \cite{li2012quantum,martone2012anisotropic,chen2017collective}. 
The phases are (i) the stripe (supersolid) phase, (ii) the plane-wave (polarized) phase, and (iii) the mixed (single-minimum) phase \cite{Chuanzhou:2016}. 

In the following, we consider the case $g > g_{12} > 0$, which yields the miscible phase and the modulationally stable condition in the absence of SOC. 
We fix the value $g=50$ and $g_{12}=0.95g$. The typical ground state solutions with and without the SOC are shown in Fig.~\ref{Fig1}. 
With these coupling constants, the phase (i) appears for $0<\Gamma \lesssim 0.2$, (ii) for $0.2 \lesssim \Gamma \lesssim 1.0$, and (iii) for $1.0 \lesssim \Gamma$. 
The transition between phase (ii) and (iii) is consistent with the property of the dispersion relation of the single particle Hamiltonian Eq.~\eqref{singleph} for $V_\text{tr} = 0$, 
which is given by 
\begin{equation}
\epsilon_{\pm} = \frac{k^2}{2} + \frac{\gamma^2}{2} \pm \sqrt{k^2 \gamma^2 + \Gamma^2} \label{sdispersion};
\end{equation}
the branch $\epsilon_{-}$ has a change between single- and double-minimum structure at $\Gamma = \gamma$ ($\gamma=1$ in our unit).
We choose the values of the Rabi coupling $\Gamma =$0.1, 0.5, and 1.5, corresponding to three different phases, to study the nonlinear dynamics caused by MI for the initial state in Fig.~\ref{Fig1}(a). 

\subsection{MI in BECs with the Raman-induced SOC}
\begin{figure}[!htbp] 
  \includegraphics[width=0.8\linewidth,scale=1]{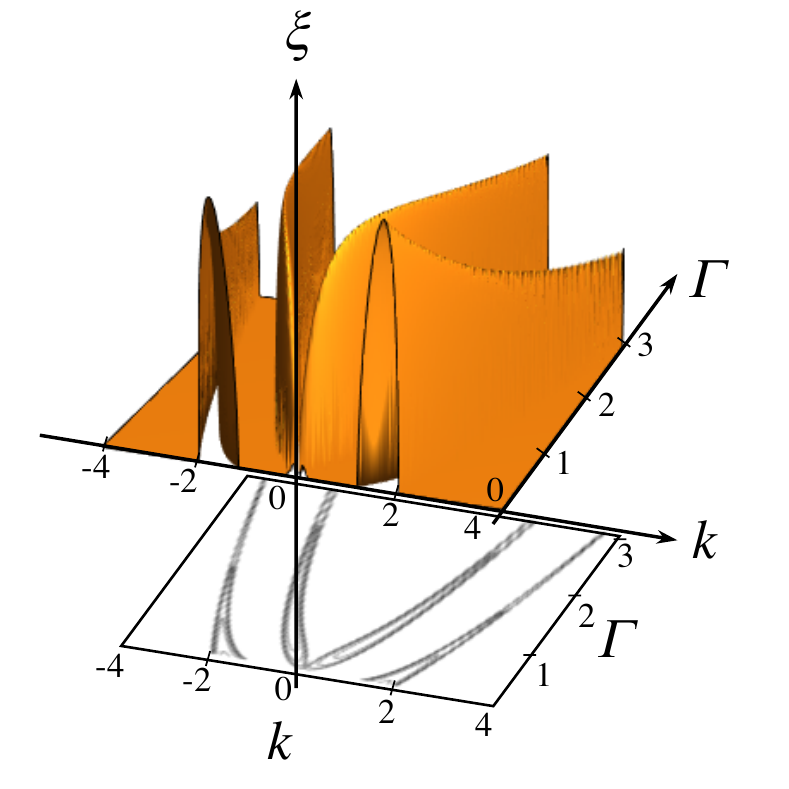}
 \caption{\label{FigMIcond}{\footnotesize (Color online) MI gain defined as $\xi = \text{Im}(\Omega)$ with the excitation frequency $\Omega$
in the $k$-$\Gamma$ plane for $g=50$, $g_{12}=0.95g$ and $n_0 = 0.0066$, which is taken from the Thomas-Fermi density at $x=0$ [see Fig.~\ref{Fig1}(a)]. 
There are four instability bands, where the two inner bands are given by the two branches $\text{Im}(\Omega_{+})$ and $\text{Im}(\Omega_{-})$ with an equivalent contribution, 
while the outer two bands are given by $\text{Im}(\Omega_{-})$. 
}}
\end{figure}

Here, we address the MI condition for the two-component BECs with the Raman-induced SOC. 
Bhat \textit{et al.} considered the MI of the miscible cw (uniform) state of two-component BECs with respect to the general parameter sets 
of the Raman-induced SOC and the interaction strengths \cite{Bhat:2015}. 
Irrespective of the combinations of the interaction strengths, the cw states are always affected by the MI 
in the presence of the SOC. In this work, we confine ourselves to the case $g > g_{12}$, $\gamma =1$ and $\Gamma > 0$. 

The condition of MI is given by the appearance of the imaginary component in the excitation frequency $\Omega$ 
for the small-amplitude modulation around the initial cw state. 
The details of this condition are given in Ref.~\cite{Bhat:2015} and the results are summarized in the Appendix~\ref{append2}. 
Figure~\ref{FigMIcond} represents the MI gain defined by $\xi = |\text{Im}(\Omega_{\pm})|$ with respect to the 
wave number $k$ and the Rabi frequency $\Gamma$. 
Here, the dispersion relation has two branches, corresponding to the (in-phase) density wave excitation $\Omega_{+}$ and 
the (out-of-phase) spin wave excitation $\Omega_{-}$. From Fig.~\ref{FigMIcond}, we see that the cw state is always dynamically unstable when 
there is the SOC. 
There are four unstable domains in the $k$-$\Gamma$ space; the two domains in the smaller-$|k|$ region, referred to as a ``primary MI band'', come from 
the two branches $\Omega_{\pm}$ with a equivalent contribution, while the other two domains, referred to as a ``secondary MI band'', in the larger-$|k|$ region 
come from only the branch $\Omega_{-}$. Thus, the SOC brings about a new regime of the MI for the two-component BECs 
which is dynamically stable without the SOC. 
%  %
% \begin{figure}[!htbp] 
%  \includegraphics[width=4.2cm,scale=1]{ground1g}
%  \includegraphics[width=4.2cm,scale=1]{ground2g}\\
%  \includegraphics[width=4.2cm,scale=1]{ground3g}%\hspace{-0.8em}
% %    \vspace{-1.0cm}
% \caption{\label{Fig2}{\footnotesize (Color online) Density profiles of ground state of Eqs. (\ref{tdgp1}) and (\ref{tdgp2}) for a) $\Gamma=0.1$, b) $\Gamma=0.5$, and c) $\Gamma=1.5$. Simulation is done for the range [-100:100] with 2048 grid points and $g_{12}=1.05g$.}}
%  \end{figure}
%%
   %

 \section{Nonlinear dynamics}\label{nonlinearMI}
In this section, we discuss the MI-induced nonlinear dynamics of the BECs with the synthetic SOC.  
We solve Eqs.~(\ref{tdgp1}) and  (\ref{tdgp2}) numerically by the split-step Fast-Fourier method; 
Details of the numerical method are described in the Appendix \ref{append1}. To generate MI, we turn on the SOC 
in the initially miscible condensates in Fig.~\ref{Fig1}(a) by introducing $\Gamma$ and $\gamma (= 1)$ suddenly at $t=0$.

The developed MI dynamics for different strengths of $\Gamma$, which gives the three different ground states in Fig.~\ref{Fig1}(b)-(d), 
are as follows.  

%----------------------------------- 
\subsubsection{$\Gamma=0.1$}
%----------------------------------
 \begin{figure*}[!htbp] 
      % \vspace{0pt}
        \includegraphics[width=15.0cm,scale=1]{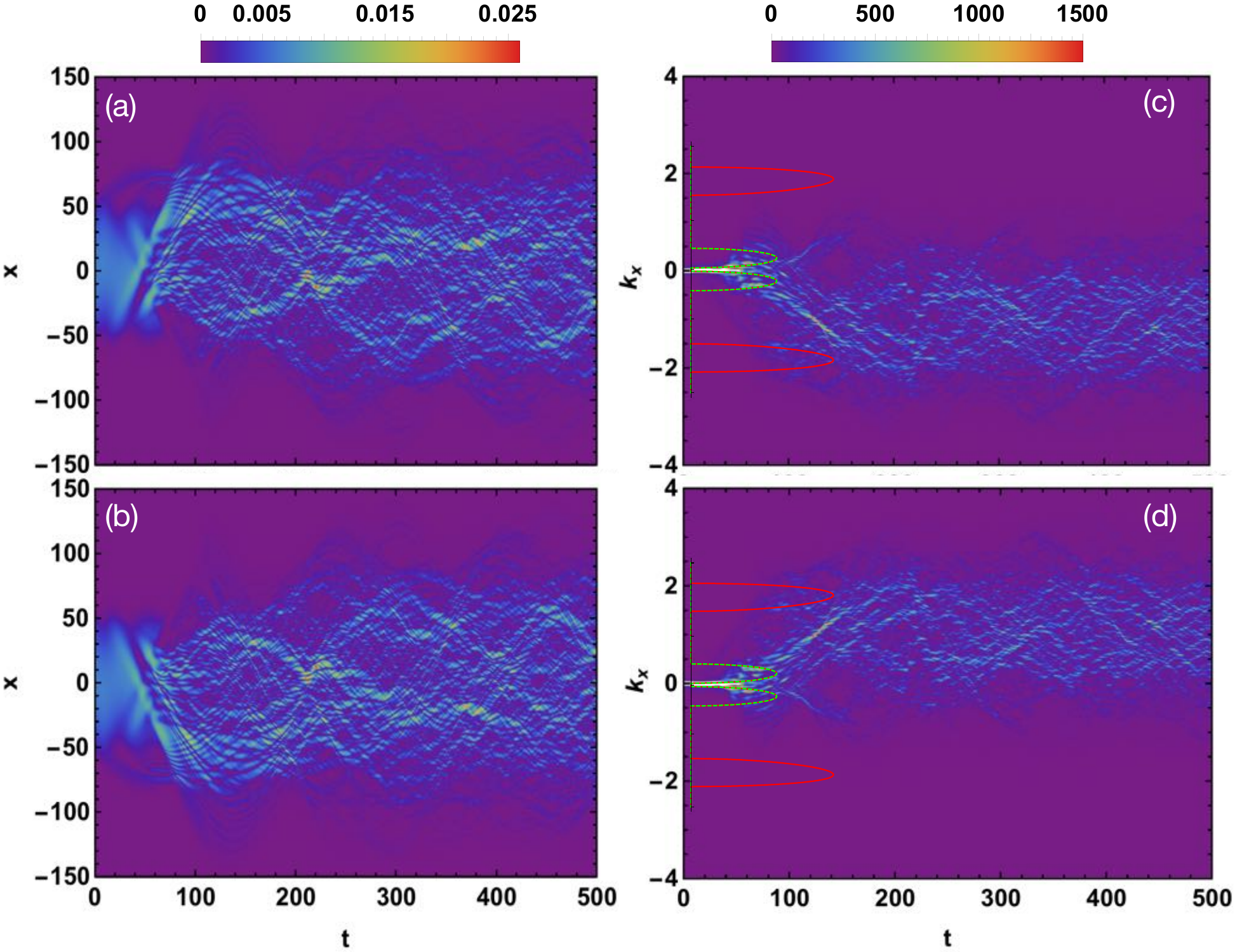} \\
        \includegraphics[width=17.0cm,scale=1]{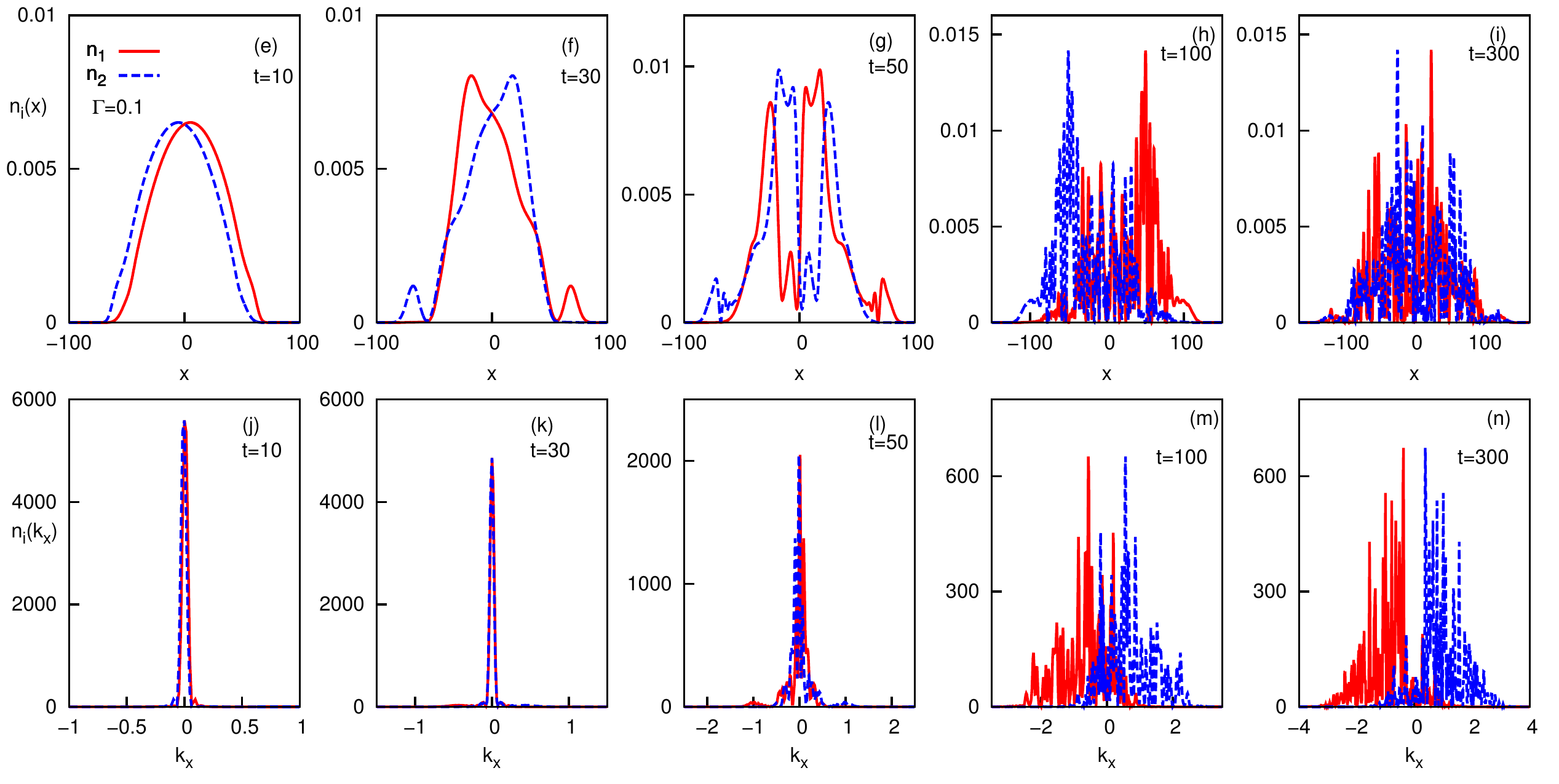}
     \vspace{-2.0pt}
 \caption{\label{Figa0.1}{\footnotesize (Color online) Time development of the condensate densities in the real coordinate space, (a) $n_1=|\psi_1(x)|^2$ 
 and (b) $n_2=|\psi_2(x)|^2$, and in the quasi-momentum space, (c) $n_1 = |\phi_1(k)|^2$ and (d) $n_2=|\phi_2(k)|^2$ for $\Gamma=0.1$. The solid red and the dashed green curves represent the MI gain $\xi$ [Eq.~\eqref{eq. gain} and the cross section of Fig.~\ref{FigMIcond} at $\Gamma=0.1$] for $\Omega_{-}$ and $\Omega_{+}$, respectively. 
The lower panels show the snapshots of the density profiles at different times: (e)-(i) in the coordinate space and (j)-(n) in the quasi-momentum space. 
Simulation is done for  the range [-200:200] with 4096 grid points in the coordinate space.}}
%      \vspace{-15pt}
%\label{1}
  \end{figure*}
%---------------------------------------

Figure~\ref{Figa0.1} shows the time development of the MI-induced spatial pattern formation from the cw state for $\Gamma = 0.1$. 
The upper left panels (a, b) and the lower panels (e-i) represent the dynamics of the condensate densities $n_i (x)= |\psi_i(x)|^2$ 
in the coordinate space. The motions of both components behave similarly and keep inversion symmetry with respect to $x=0$. 
First, the condensates make out-of-phase dipole motions to shift the centers of mass from each other. 
As time evolves, the densities of both components are well separated and break into smaller domains. After $t=50$, 
the both components fragment into the non-periodic short-wavelength domains and continue to make a chaotic oscillation. 

Further insight can be seen in the dynamics of the wave function in the quasi-momentum space.  
The upper left panels (c, d) and the lower panels (j-n) represent the dynamics of the densities $n_i(k_x) = |\phi_i(k_x)|^2$ 
of the Fourier component $\phi_i(k_x) = \int dx \psi_i(x) e^{-ik_x x}$, where in a lab frame Fig.~\ref{Figa0.1}(j) corresponds to the initially overlapped wave packets sitting at trap center with left-going spin-up momentum $-\hbar k_R$ and a right-going spin-down momentum $ \hbar k_R$. 
The panels (c, d) also show the MI gain (cross section of Fig.~\ref{FigMIcond} at $\Gamma=0.1$) for clarity. 
In the quasi-momentum space, the SOC $\pm i \gamma \partial_x$ in Eqs.~\eqref{tdgp1} and \eqref{tdgp2} contributes 
as a linear potential $+\gamma k_x$ and $-\gamma k_x$ for $\phi_1$ and $\phi_2$, respectively. 
Thus, $\phi_1$ and $\phi_2$ are basically driven to the negative and positive direction in the $k$-space, respectively. 
First, since the primary MI bands exist around $k=0$, the low-energy dipole motions grow spontaneously due to the MI. 
This dipole motion induces the spin-wave excitations and leads to the generation of unstable modes in the secondary bands. 
The panels (c) and (d) clearly show this transfer process of the wave component from small- to large-$k$ region, where the $n_{1(2)}(k_x)$ propagates to the 
negative (positive) wave number of the secondary MI band associated with $\Omega_{-}$. During the time evolution, from the panels (j-n), we can see that the waves collapse into the complicated short-wavelength domains but 
show clear phase separation in the positive and negative range of the quasi-momentum space because of the constant bias of the SOC. 
 Since the double minima of Eq.~\eqref{sdispersion} exist at $k = \pm \sqrt{\gamma^2 - \Gamma^2}/\gamma \approx \pm 1$, 
$|\phi_{1}(k_x)|^2$ and $|\phi_{2}(k_x)|^2$ eventually distribute around $k_x \sim 1$ on average.

 \subsubsection{$\Gamma=0.5$}
    %------------------------------
     \begin{figure*}[!htbp] 
      % \vspace{0pt}
       \includegraphics[width=15.0cm,scale=1]{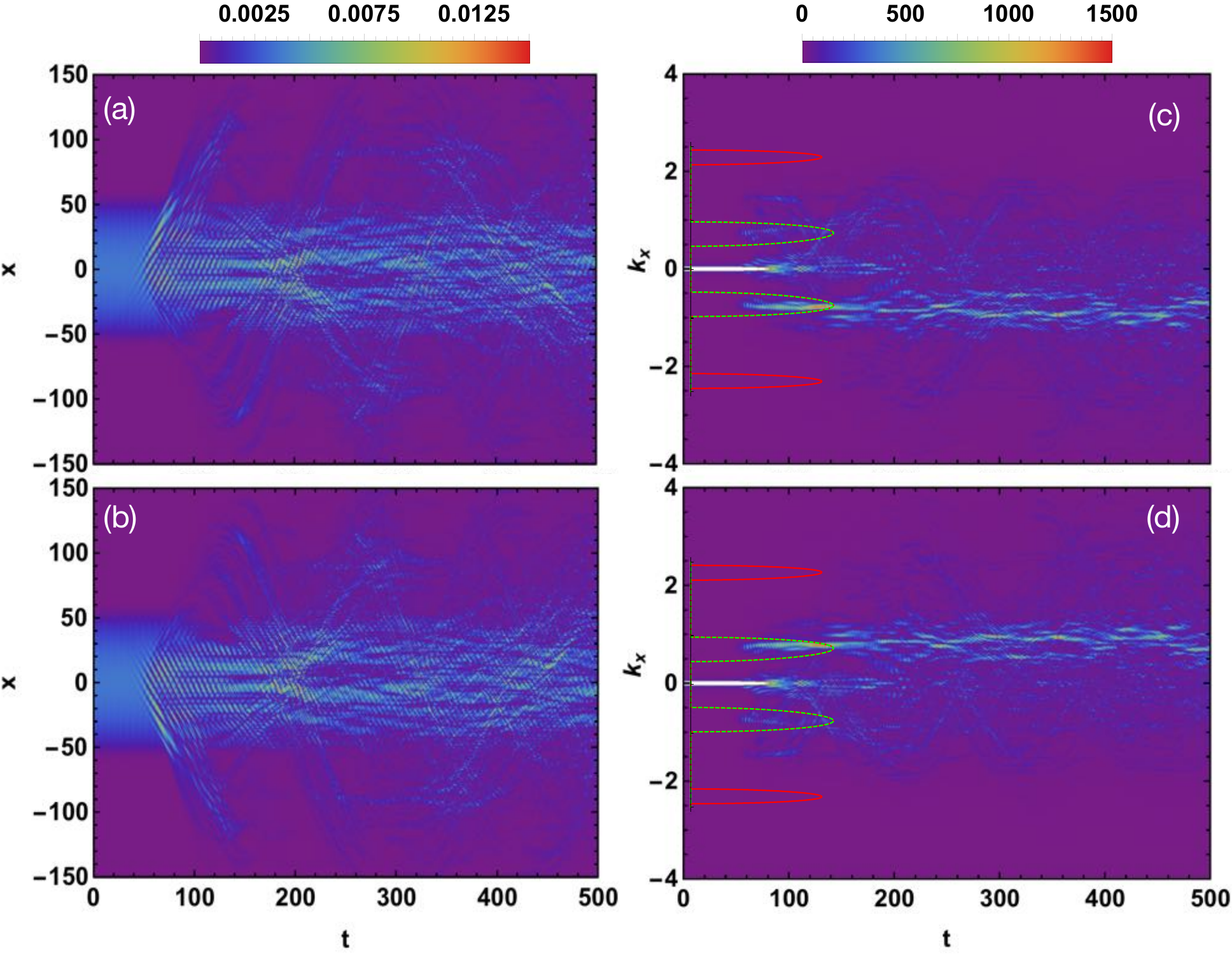} \\
       \includegraphics[width=17.0cm,scale=1]{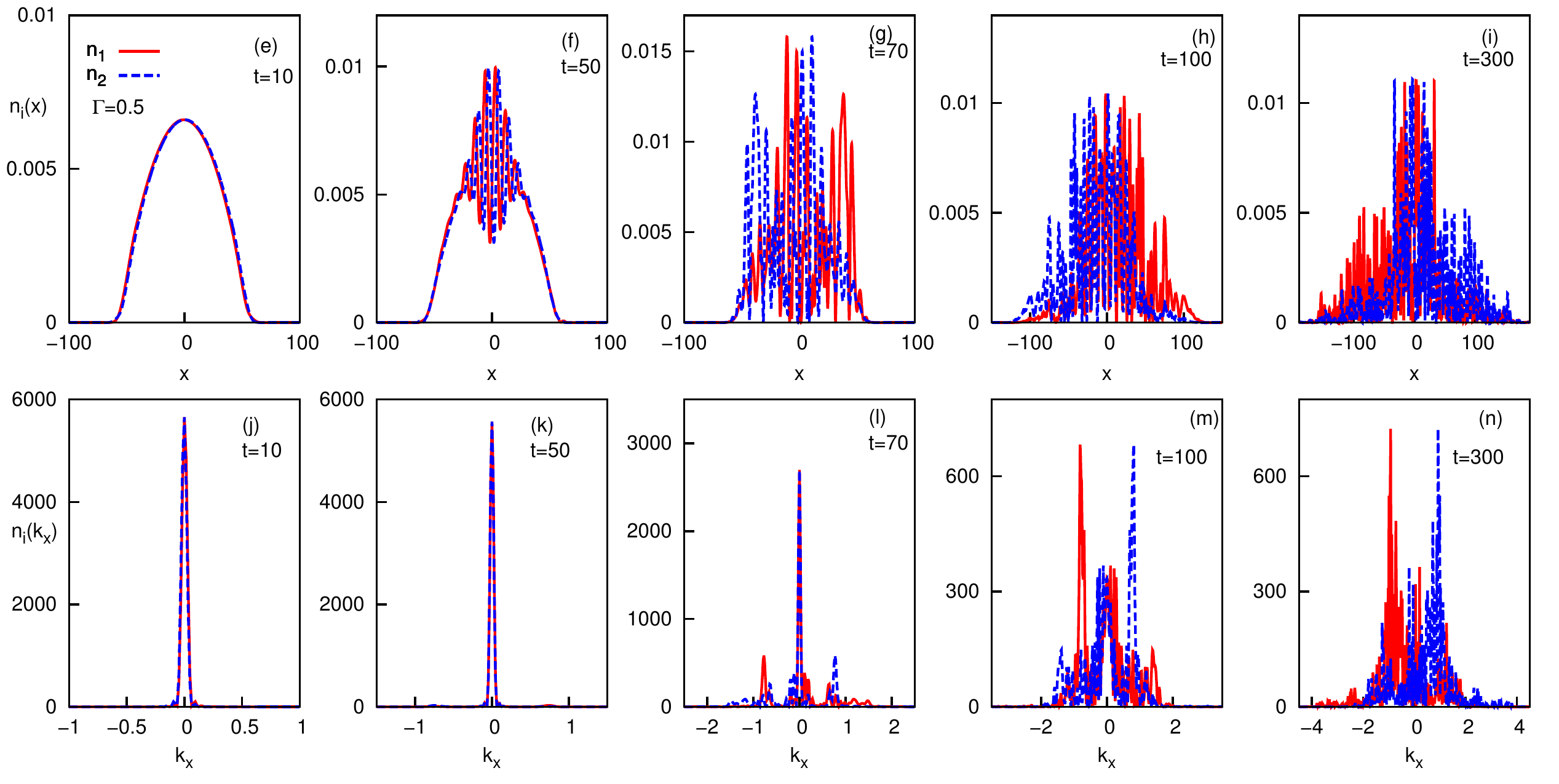}
     \vspace{-2.0pt}
 \caption{\label{Figa0.5}{\footnotesize (Color online) The results similar to Fig.~\ref{Figa0.1} for $\Gamma=0.5$.}}
  \end{figure*}
For the increased value of Rabi frequency, $\Gamma=0.5$, the evolution of the density in the coordinate space is shown in Fig.~\ref{Figa0.5}(a,b) and snap shots of density at different times are depicted in Fig.~\ref{Figa0.5}(e-i). 
Initially a few density stripes appear in both components at the center of the condensate and grow in-phase. As time evolves, the number of stripes increases and outer density exhibits large oscillatory behaviors. Also, one can see that the initial shift of the center-of-mass shift observed for $\Gamma =0.1$ is suppressed here.  
Finally, the both components are fragmenting into the non-periodic short-wavelength domains as in the case of $\Gamma =0.1$. 

In the Fourier space, Figs.~\ref{Figa0.5}(c,d) and (j-n) show that the initially excited wave vector exactly matches with the analytically predicted $k_\text{max}$ of the primary MI band, as seen 
%by the merged MI gain $\xi$ of the growing modes $\Omega_{-}$ and $\Omega_{+}$ 
in Fig.~\ref{Figa0.5}(c,d). Although a small amount of excitations appears at the secondary bands at later times, the increased distance (when compared to the that of $\Gamma=0.1$ case) between the primary the secondary bands keeps the excitations well in the primary bands. 
Also, the dynamics of the components ($\phi_1$ and $\phi_2$) are not completely separated in to the positive and negative wave vector values as in the case of $\Gamma=0.1$. This is expected because of the reduction in the effective potential given by the combination of Rabi coupling and the SOC. These properties are again 
consistent with the single-particle dispersion of $\epsilon_-$, where the separation of the double minima are reduced compared to the case of $\Gamma= 0.1$.

  %---------------------  %

 \subsubsection{$\Gamma=1.5$}
   \begin{figure*}[!htbp] 
      % \vspace{0pt}
     \includegraphics[width=15.0cm,scale=1]{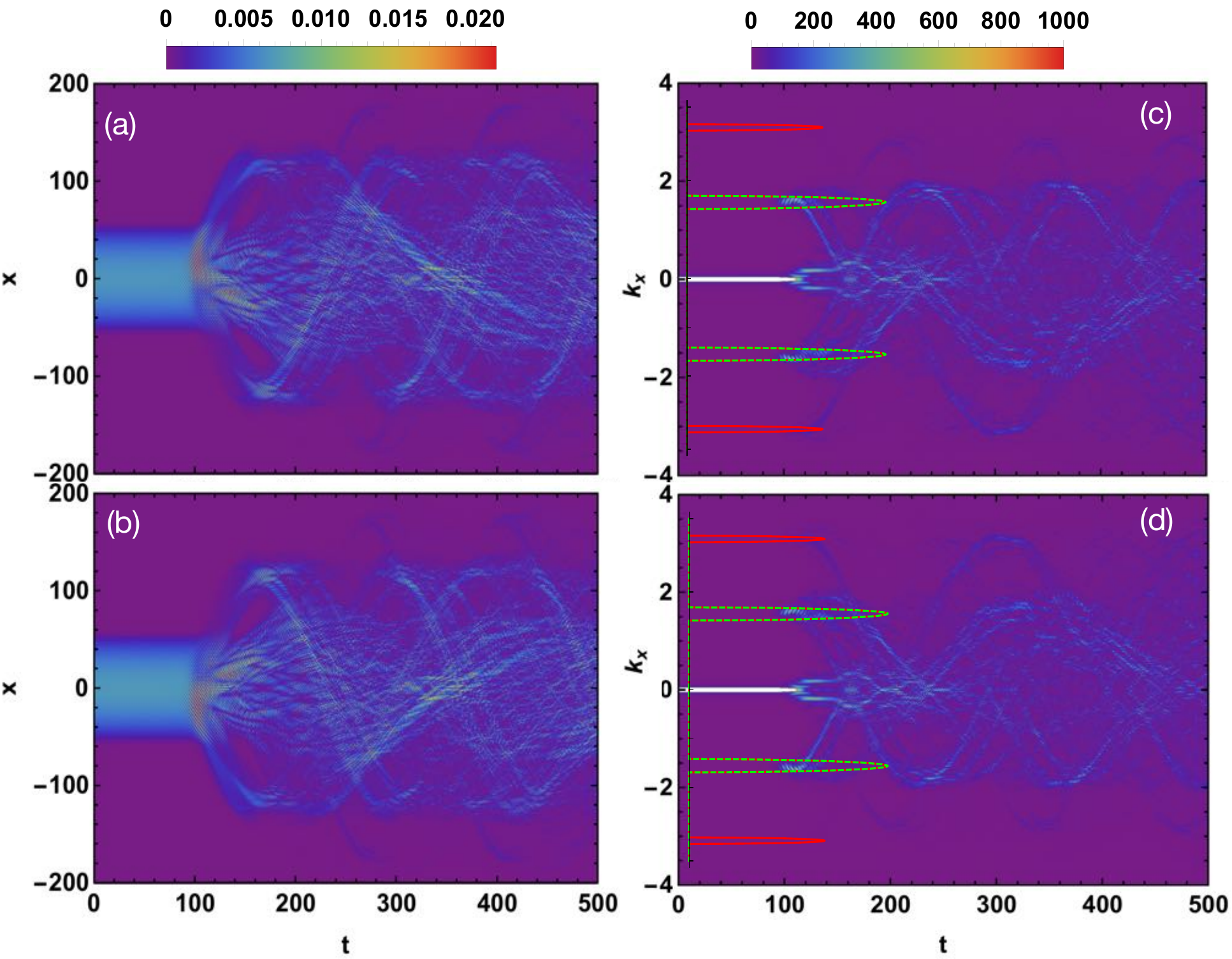} \\
    \includegraphics[width=17.0cm,scale=1]{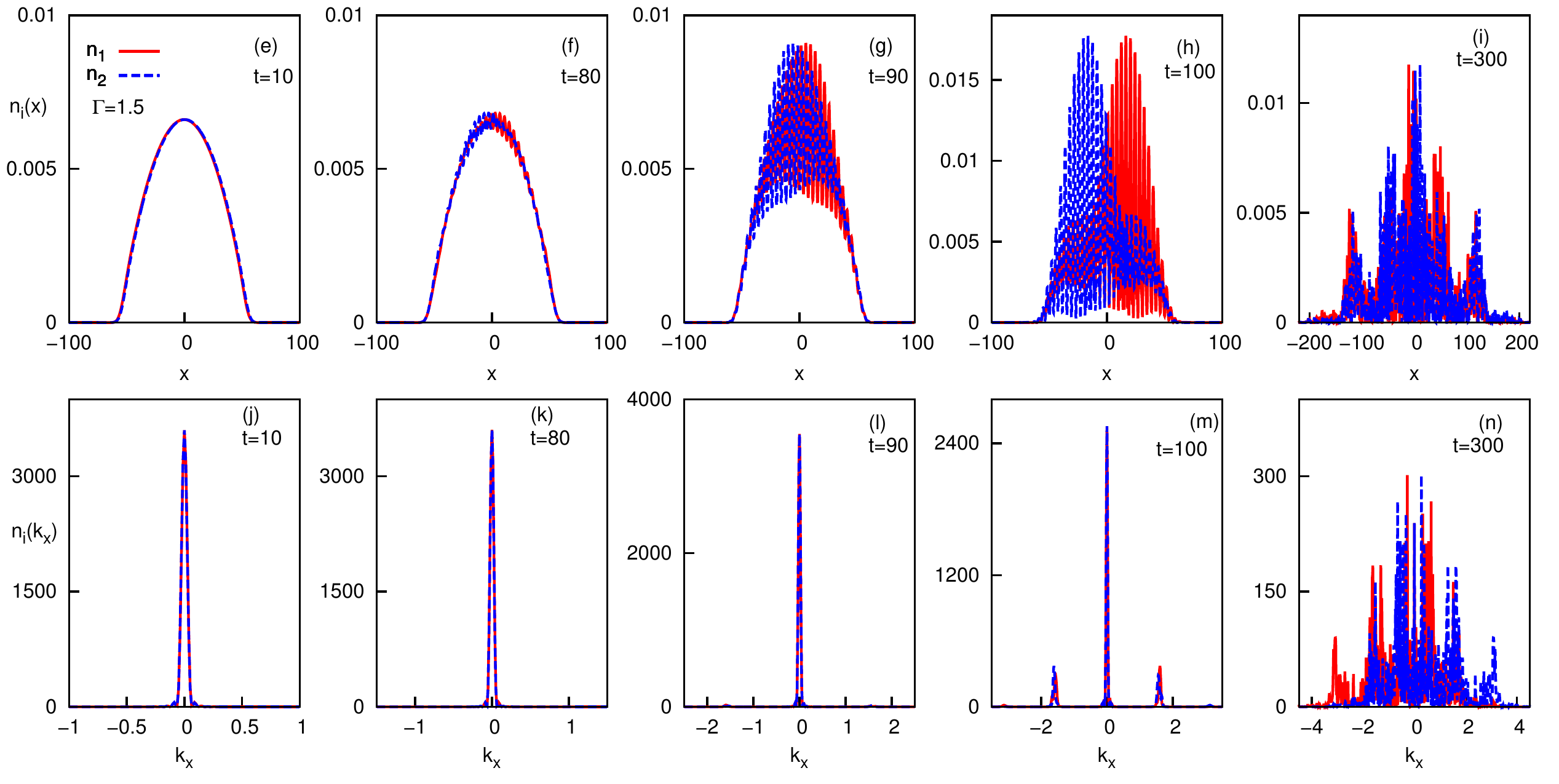}
     \vspace{-2.0pt}
 \caption{\label{Figa1.5}{\footnotesize (Color online) The results similar to Fig.~\ref{Figa0.1} for $\Gamma=1.5$.}}
  \end{figure*}
    %---------------------------------------
       \begin{figure}[!htbp] 
      % \vspace{0pt}
     \includegraphics[width=8.6cm,scale=1]{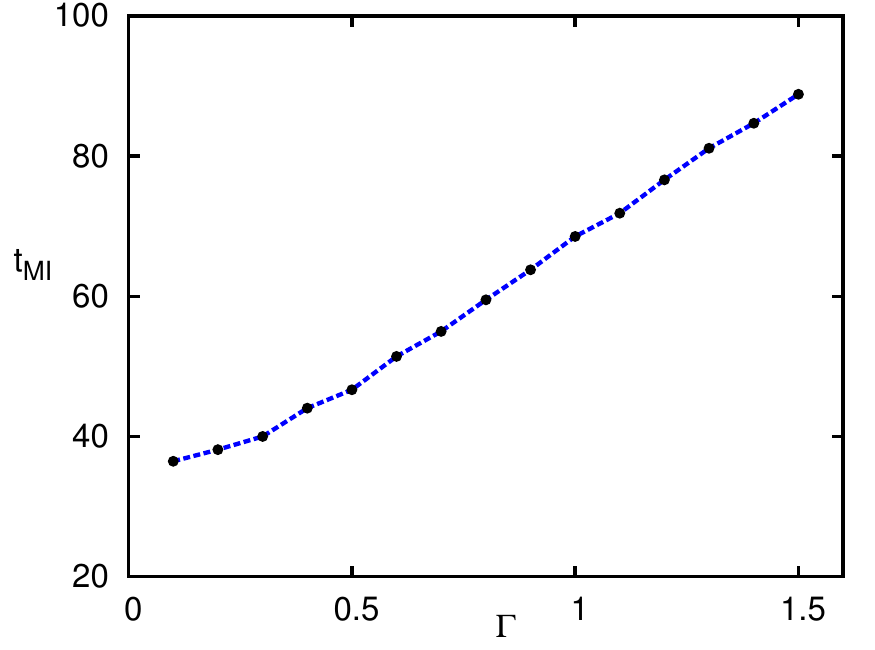} \\
     \vspace{-2.0pt}
 \caption{\label{Fig_MI_time}{\footnotesize (Color online) The modulation instability time (black dot) is plotted for varying Rabi frequency $\Gamma$.  The dashed line guides the eye.}}
  \end{figure}
    %---------------------------------------

Fig.~\ref{Figa1.5}(a,b)  and  Fig.~\ref{Figa1.5}(e-i) show that the spatial pattern formation for $\Gamma=1.5$ which resemble that of $\Gamma=0.5$. The density modulations of the both components generate spontaneously at the center, growing to large amplitude density waves. Then, each component makes a spatial separation [Fig.~\ref{Figa1.5}(h)] and continues a complicated chaotic oscillation. In the Fourier space, Fig.~\ref{Figa1.5}(c,d) and Fig.~\ref{Figa1.5}(j-n) display that initial modulation appears at the primary and secondary bands, consistent with the MI analysis. Here, the excitations happens symmetric to $k = 0$ in the primary bands, meanwhile, in the secondary band excitations are asymmetric due to the increased value of $|k|$. 
Since the dispersion $\epsilon_{-}$ of Eq.~\eqref{sdispersion} has a single minimum at $k=0$ for $\Gamma=1.5$, the separation of $\phi_i$ in the quasi-momentum space does not occur; the fragmented domains of $\phi_i$ oscillate around $k=0$ as seen in Fig.~\ref{Figa1.5}(c,d). 

Fig.~\ref{Fig_MI_time} shows the plot of the critical time $t_{\text{MI}}$ at which MI induced chaotic dynamics starts. It is found that with Rabi frequency $\Gamma$,  $t_{\text{MI}}$  also increases and exhibits a linear relation for $\Gamma \gtrsim 0.5$. 
This is because the primary MI band are well separated from the $k \simeq 0$ region, thus, there require some time to reach the finite wave length modes 
through the nonlinear mode couplings. If we input the initial noise corresponding to the wave number at the primary or secondary MI bands, the MI grows quickly after  the SOC is turned on. 

\section{Conclusions} \label{conclusion}
In conclusion, we investigate the nonlinear dynamics induced by MI in two-component BECs with Raman-induced synthetic SOC. 
In the previous studies, the MI was predicted to occur for arbitrary choices of the intra- and intercomponent coupling constants. 
We demonstrated that even for miscible two-component BECs, which is dynamically stable without SOC, the MI 
can take place and cause complicated nonlinear dynamics of pattern formation. 
The onset of the nonlinear evolution is consistent with the theoretical prediction of the MI for homogeneous 
two-component BECs. The presence of the primary and secondary MI bands induces the characteristic two-step 
nonlinear evolution of the pattern formation. At the later stages of the evolutions, the wave functions in the 
quasi-momentum space undergo the separation due to the asymmetric feature of the SOC, depending on the values of the 
Rabi coupling corresponding to the three ground state phases.

Although the MI yields the very complicated dynamics as shown in Figs.~\ref{Figa0.1}, \ref{Figa0.5} and \ref{Figa1.5}, the ground states 
are well-defined shape as in Figs.~\ref{Fig1}(b-d). It is interesting to see the relaxation process from the 
strongly nonequilibrium fragmented states to the ordered ground states by introducing some dissipation 
or fluctuation effects. Also, revealing nonlinear dynamics in higher-dimensions is an interesting direction for future studies. 
We hope that this work further stimulates the studies of the nonlinear dynamics in multicomponent BECs.

\section{Acknowledgements}
T. M acknowledges financial support from IBS (Project Code No. IBS-R024-D1).  The work of K.K. is partly supported by KAKENHI from the Japan Society for the Promotion of Science (JSPS) Grant-in-Aid for Scientific Research (KAKENHI Grant No. 18K03472). 

%---------------------------------------------------
%\clearpage
%\newpage
\appendix
\section{Analytical formulation}\label{append2}
In Ref. \cite{Bhat:2015}, the condition of MI under the Eqs.~\eqref{tdgp1} and \eqref{tdgp2} has been derived analytically. 
In a uniform system $V_\text{tr} = 0$, the wave functions are expanded from the uniform density as 
$\psi_j = (\sqrt{n_{j0}} + \delta \psi_j)$ and the linear stability analysis for the fluctuation $\delta \psi_j$ 
yields the eigenfrequecy 
 \begin{equation}
{\Omega _{\pm }^{2}}=\frac{1}{2}[\Lambda \pm \sqrt{\Lambda ^{2}+4R}],
\label{Omega}
\end{equation}%
\newline
where%
\begin{equation}
\Lambda =2k_x^{2}\gamma ^{2}+\frac{1}{2}{(k_x^{2}-2\Gamma )(k_x^{2}+G)}+2\Gamma
G_{12},
\end{equation}%
\begin{equation}
R=(S-\Lambda _{+}\Lambda _{-}),\newline
\end{equation}%
\begin{equation}
\Lambda _{\pm }=\frac{\Lambda \pm \sqrt{2\Lambda _{1}(\Lambda -\Lambda _{1})}%
}{2},
\end{equation}%
\begin{gather}
S=\frac{1}{4}\left\{ {(k_x^{2}-2\Gamma )^{2}}G_{12}^{2}+(G+k_x^{2})^{2}\Gamma ^{2}\right\} \notag \\ +\frac{1}{2}\left\{\gamma ^{2}k_x^{2}
(G +k_x^{2})(k_x^{2}-2\Gamma) \right\} ,
\end{gather}%
\begin{equation}
\Lambda _{1}=\frac{1}{2}{(k_x^{2}-2\Gamma )(G+k_x^{2})},
\end{equation}%
and
\begin{equation}
\begin{split}
G_{1}&\equiv 4gn_0-2\Gamma ,~G_{2}\equiv 4gn_0-2\Gamma,~G=G_{1}=G_2 ,\\~G_{12}&\equiv
2g_{12}n_{0}+\Gamma,\newline
\label{G}
\end{split}
\end{equation}
where $n_{10}=n_{20}=n_{0}=|\psi_1|^2=|\psi_2|^2$. 

The MI gain is defined as
\begin{equation}
 \xi=|Im(\Omega_{\pm})|.
 \label{eq. gain}
\end{equation}

 \section{Numerical Simulation}\label{append1}
In our simulations, we employed the split-step Fast-Fourier method to solve the GP equations. 
The total Hamiltonian of the BEC with SOC is
\begin{equation}
   \begin{split}
\mathcal{H} = \mathcal{H}_1+\mathcal{H}_2+\mathcal{H}_3, 
   \label{eq:hamil}
        \end{split}
\end{equation}
where
$$
\mathcal{H}_1 = \int dx \biggl[ \frac{1}{2}\biggl(\left|\frac{\partial \psi_1}{\partial x}\right|^2+\left|\frac{\partial \psi_2}{\partial x} \right|^2 \biggr)+i \gamma \biggl(-\psi^{\ast}_1\frac{\partial \psi_1}{\partial x}+\psi^{\ast}_2\frac{\partial \psi_2}{\partial x}\biggr)\biggr],
$$
$$\mathcal{H}_2 =\int dx \left[ V(x)n+\frac{1}{2}g n_1^2+\frac{1}{2}g n_2^2+g_{12}n_2 n_1\right],
$$
$$\mathcal{H}_3=\int dx \Gamma \left(\psi^{\ast}_1 \psi_2+\psi^{\ast}_2 \psi_1 \right),
$$
with the densities  $n_1=|\psi_1|^2$, $n_2=|\psi_2|^2$, and $n=|\psi_1|^2+|\psi_2|^2$. 
Here, Hamiltonian is written in the dimensionless form. 
In the time development described by Eqs.~(\ref{tdgp1}) and (\ref{tdgp2}), 
the Hamiltonian is split into three integrable parts as $\mathcal{H}_1$, $\mathcal{H}_2$ and $\mathcal{H}_3$, 
\begin{equation}
\begin{split}
&L_{\mathcal{H}_1}:\ 
\left\{
\begin{array}{rl}
i\frac{\partial \psi_1}{\partial t} =\big[ - \frac{1}{2}\frac{\partial^2}{\partial x^2}-i\gamma \frac{\partial}{\partial x}\big]\psi_1\\
i\frac{\partial \psi_2}{\partial t} =\big[ - \frac{1}{2}\frac{\partial^2}{\partial x^2}+i\gamma\frac{\partial}{\partial x}\big]\psi_2
\end{array} \right.
\end{split}
\label{eq:LH1}
\end{equation}
 \begin{equation}
\begin{split}
&L_{\mathcal{H}_2}:\ 
\left\{
\begin{array}{rl}
i\frac{\partial \psi_1}{\partial t} =\big[ V(x)+g|\psi_1|^2+g_{12}|\psi_2|^2 \big]\psi_1\\
i\frac{\partial \psi_2}{\partial t} =\big[ V(x)+g|\psi_1|^2+g_{12}|\psi_2|^2 \big]\psi_2
\end{array} \right.
\end{split}
\label{eq:LH2}
\end{equation}
and
 \begin{equation}
\begin{split}
&L_{\mathcal{H}_3}:\ 
\left\{
\begin{array}{rl}
i\frac{\partial \psi_1}{\partial t} =\Gamma \psi_2\\
i\frac{\partial \psi_2}{\partial t} =\Gamma \psi_1
\end{array} \right.
\end{split}
\label{eq:LH3}
\end{equation}

The evolution of corresponding resolvent operators with the time step $\tau$ is written as 
 \begin{equation}
\begin{split}
&e^{\tau\mathcal{H}_1}:\ 
\left\{
\begin{array}{rl}
\phi_1(k_x,t+\tau)=\phi_1(k_x,t) e^{-i \tau(\frac{1}{2}k_x^2+ \gamma k_x)} \\
\phi_2(k_x,t+\tau) =\phi_2(k_x,t) e^{-i \tau(\frac{1}{2}k_x^2- \gamma k_x)},
\end{array} \right.
\end{split}
\label{eq:LH1}
\end{equation}
where  Fourier-Transformed term
$$\phi_i(k_x,t)=\int_{-\infty}^{\infty}\psi_i(x,t)e^{-ik_x x}dx,$$ and 
its  inverse transformed term 
$$\psi_i(x,t+\tau)=\frac{1}{2\pi}\int_{-\infty}^{\infty}\phi_i(k_x,t+\tau)e^{ik_x x}dk_x,$$.
 \begin{equation}
\begin{split}
&e^{\tau\mathcal{H}_2}:\ 
\left\{
\begin{array}{rl}
\psi_1(x,t+\tau)=\psi_1(x,t) e^{-i \tau(V(x)+g_1|\psi_1|^2+g_{12}|\psi_2|^2)}\\
\psi_2(x,t+\tau) =\psi_2(x,t) e^{-i \tau(V(x)+g_2|\psi_1|^2+g_{12}|\psi_1|^2)}
\end{array} \right.
\end{split}
\label{eq:LH2}
\end{equation}
 \begin{equation}
\begin{split}
&e^{\tau\mathcal{H}_3}:\ 
\left\{
\begin{array}{rl}
\psi_1(x,t+\tau)= \cos(\Gamma \tau)\psi_1(x,t)-i\sin(\Gamma \tau)\psi_2(x,t)\\
\psi_2(x,t+\tau) =\cos(\Gamma \tau)\psi_2(x,t)-i\sin(\Gamma \tau)\psi_1(x,t).
\end{array} \right.
\end{split}
\label{eq:LH3}
\end{equation}

Now the evolution of the total Hamiltonian, $\mathcal{H}$ is 
 \begin{equation}
\begin{split}
e^{\tau\mathcal{H}}= e^{\frac{1}{2}\tau\mathcal{H}_1}e^{\frac{1}{2}\tau\mathcal{H}_3}e^{\tau\mathcal{H}_2}e^{\frac{1}{2}\tau\mathcal{H}_3}e^{\frac{1}{2}\tau\mathcal{H}_1}
\end{split}
\label{eq:LH}
\end{equation}
In the simulation we fix $g=50 $,  $g_{12}=0.95g$, $\gamma=1$, and $\tau=0.005$.

      %-----------------------------------------
%
\bibliographystyle{apsrev4}
\let\itshape\upshape
%\normalem
\bibliography{reference}
\end{document}